\documentclass[useAMS,usenatbib,fleqn]{mn2e}
\usepackage[latin1]{inputenc}
\usepackage{amsmath}
\usepackage{amsfonts}
\usepackage{amssymb}
\usepackage{setspace}
\usepackage{epsfig}
\usepackage{lscape,graphicx}
\usepackage{natbib}
\usepackage{textcomp}
\usepackage{epstopdf}
\usepackage{graphicx}
\usepackage{caption}
\usepackage{float}
\usepackage{aas_macros}
\usepackage{color}
\usepackage{ulem}

\def\lsim{\mathrel{\lower0.6ex\hbox{$\buildrel {\textstyle <}
 \over {\scriptstyle \sim}$}}}
\def\gsim{\mathrel{\lower0.6ex\hbox{$\buildrel {\textstyle >}
 \over {\scriptstyle \sim}$}}}

\def \d {{\rm d}}

\def \rvir{$  R_{\rm vir}$}
\def \mvir {  M_{\rm vir}}
\def \Mvir {$  M_{\rm vir}$}

\newcommand{\hmsun}{{\,\rm h^{-1}M}_\odot}
\newcommand{\hmpc}{{\,\rm \textit{h}^{-1}Mpc}}

\begin{document}

\title[Abundance and environment of haloes]{The abundance and  environment of dark matter haloes}

\author[Metuki et al.] 
{\parbox[t]\textwidth{ Ofer Metuki$^1$, Noam I. Libeskind$^{2}$, Yehuda Hoffman$^1$}
\vspace*{6pt} \\ 
 $^{1}$Racah Institute of Physics, Hebrew University, Jerusalem 91904, 
 Israel\\ 
 $^{2}$Leibniz-Institut f\"ur Astrophysik Potsdam (AIP), An der Sternwarte 16, D--14482 
 Potsdam, Germany\\
}

\date{\today}  

\maketitle

\begin{abstract}  
 
 An open question in cosmology and the theory of structure formation is to what extent does environment affect the properties of galaxies and haloes. The present paper aims at shedding light on this problem. The paper focuses on the analysis of a dark matter only simulation and it addresses the issue of how the environment affects the abundance of haloes, which are are assigned four attributes: their virial mass, an ambient density calculated with an aperture that scales with $R_{vir}$ ($\Delta_M$), a fixed-aperture ($\Delta_R$) ambient density, and a cosmic web classification (i.e. voids, sheets, filaments, and knots, as defined by the V--web algorithm). $\Delta_M$ is the mean density around a halo evaluated within a sphere of a radius of $5$\rvir, where \rvir\  is the virial radius. $\Delta_R$ is the density field Gaussian smoothed with $R=4\hmpc$, evaluated at the center of the halo.
  The main result of the paper is that the difference between haloes in different web elements stems from the difference in their mass functions, and does not depend on their adaptive-aperture ambient density. A dependence on the fixed-aperture ambient density is induced by the cross correlation between the mass of a halo and its fixed-aperture ambient density.

\end{abstract}


%%%%%%%%%%%%%%%%%%%%%%%%%%%%%%%%%%%%%%%%%%%%%%
%%%%%%%%%%%%%%%%%%%%%%%%%%%%%%%%%%%%%%%%%%%%%%
\section{Introduction}  
\label{sec:intro}

Galaxies know about their neighbourhood. Namely, there are correlations between a galaxy's properties and its environment. Arguably, the most striking correlation is the one between the morphological type of a galaxy and its ambient density \citep{1980ApJ...236..351D}, which a theory of galaxy formation ought to be able to explain. The morphology--density relation provides an important clue and insight into a comprehensive theory of galaxy formation. 
A number of models exist to explain the environmental dependencies of galaxy properties. Many of these suggest that environment can regulate the gas supply available to galaxies for star formation; processes such as harassment \citep[wherein galaxy morphology is transformed due to frequent high speed encounters; e.g.][]{1996Natur.379..613M}, strangulation  \citep[wherein the hot gas supply is slowly removed, thereby gradually reducing a galaxy's star formation rate; e.g. ][]{2000ApJ...540..113B}, stripping via e.g. ram pressure \citep{1972ApJ...176....1G}  or tidal effects \citep{1962AJ.....67..471K}, and mergers \citep[that transform spirals into ellipticals; e.g.][]{1972ApJ...178..623T},  are a number of ``quenching'' effects that are triggered by changes in a galaxy's environment \citep[e.g.][]{2008MNRAS.387...79V}. Given the uncertainties  associated with the numerical implementation of the baryonic physics at play, one may look into dark matter (DM) haloes as a crude and 'first order' proxy for galaxies and as the site of galaxy formation. Such an approach has motivated us to study the abundance of DM haloes in cosmological N--body simulations and its dependence on their environment.

The study of the halo mass function plays a major role in the development of a theory of structure and galaxy formation, starting with the seminal work by \citet{1974ApJ...187..425P}. It is by now clear that the efficiency of halo formation depends on the environment in which the halo resides. 
Dense environments are more efficient in producing massive haloes than less dense ones, resulting in a mass function that is skewed towards more massive haloes \citep{2007ApJ...654...53M,1999MNRAS.302..111L}.  This has prompted an intensive study as to how to define and quantify the notion of environment.

\citet{2012MNRAS.419.2670M} conducted a thorough comparison of 20 different algorithmic approaches for the definition of the environment. These authors correctly stated that almost all methods are based on either nearest--neighbour statistics or on an evaluation of the ambient density in fixed 2- or 3--dimensional volumes. Both approaches provide a scalar measure of the environment, namely they do not define preferred directions. However, visual inspection of the distribution of (observed) galaxies and (simulated) DM haloes gives rise to an intricate structure of vast voids, planar sheets, long filaments and dense and compact knots - the so--called 'cosmic web' \citep{1996Natur.380..603B}. The notion of the cosmic web does not only provide a link between the local density of a region and the topography of its environment, but also defines preferred directions. The sequence of web elements, ranging from voids, to sheets, to filaments, and to knots, corresponds to a sequence of: a. increasing local density \citep[e.g.][]{2012MNRAS.425.2049H}; b. increasing mass of DM haloes \citep[e.g.][]{2015MNRAS.446.1458M}; 
c. by implication, the increasing local density leads to an  increasing ratio of early to late type galaxies. An important question arising from this is what is the main mechanism behind these correlations?

In the context of the present paper we adopt a rather limited view of DM haloes, and 
characterize each halo by the following properties: its virial mass, its web environment, and the ambient density within which it resides. 
A short and  incomplete review of these properties of haloes is presented here. The abundance of haloes will be studied here in the framework of such a characterization.

\citet{2007MNRAS.381...41H} analyzed the properties of haloes with respect to the cosmic web, defined by the eigenvalues of the Hessian of the gravitational potential, and showed that the halo mass function varies with web classification. They also showed that the high mass end of the halo mass function increases as the web sequence goes from voids to knots.  \citet{2015MNRAS.447.2683A} have recently studied the dependence of the halo mass function on both the ambient density and on the cosmic web (as defined by \citealt{2009MNRAS.396.1815F}),
 and  concluded that ``...
  to a good approximation, the abundance of haloes in different environments depends only on their densities, and not on their tidal [web] structure.''  \citet{2015MNRAS.448.3665E} have investigated the dependence of galaxy luminosity on the cosmic web and local density within the GAMA survey. These authors found that the modulation of the luminosity function with environment can be accounted for by the luminosity function's dependence on density, and found ``no evidence of a direct influence
of the cosmic web on the galaxy luminosity function''. \citet{2015MNRAS.446.1458M} studied the dependence of properties of simulated galaxies in a suit of gas--dynamical cosmological simulation on the cosmic web  \citep[as defined by][]{2012MNRAS.425.2049H} and found a strong dependence of the halo and galaxy properties on the cosmic web, but showed that this dependence can be virtually accounted for by the dependence of the haloes' mass function on the web.

\citet{1996Natur.380..603B} were the first to introduce the concept of the 'cosmic web',  yet the idea that the matter distribution can be characterized  by voids, sheets (called at the time pancakes), filaments, and knots dates back to the seminal work of \citet{1970A&A.....5...84Z}. Since then many methods and algorithms have been proposed  to developing a mathematical description of the cosmic web. 

 We follow here a path that started with \citet{2007MNRAS.381...41H,2007MNRAS.375..489H} and was further extended by \citet{2009MNRAS.396.1815F} \citep[see also the NEXUS algorithm of ][]{2013MNRAS.429.1286C}. Common to all of these methods is that the web is classified according to the number of eigenvalues of the Hessian of a scalar field (such as the gravitational potential, the matter density, or the velocity potential, namely the velocity shear tensor) which are larger than some threshold. The number of eigenvalues larger than said threshold at a given point assigns a web classification - 0 (voids), 1 (sheets), 2 (filaments) and 3 (knots) - and the eigenvectors endow preferred directions. 
The tidal web \citep{2007MNRAS.375..489H,2009MNRAS.396.1815F} has been improved upon by \citet{2012MNRAS.425.2049H}, who have replaced the gravitational tidal tensor by the velocity shear tensor, and thereby improved the spatial resolution. The velocity shear tensor based web (V--web) resolves the cosmic web down to  sub--Megaparsec scales (in numerical simulations), and has been used to examine halo spins \citep{2012MNRAS.421L.137L}, halo alignment \citep{2013MNRAS.428.2489L}, subhalo accretion \citep{2014MNRAS.443.1274L} and planes of satellites \citep{2015MNRAS.452.1052L}. The V--web is adopted here as a measure of the cosmic web.

 The aim of the present paper is to study the dependence of the abundance of haloes on their mass, cosmic web environment and the ambient density. The present analysis bears some resemblance to previous work \citep{2015MNRAS.447.2683A,2015MNRAS.448.3665E} but it also differs from those studies in a fundamental way. A distinction is made here between the fixed-aperture and adaptive-aperture approaches. The V--web is calculated 	in a multi--scale fashion, in which the spatial resolution of the V--web of a given halo scales with its virial radius \citep{2014MNRAS.443.1274L,2015MNRAS.446.1458M}. 
The ambient density is calculated here in both fashions - an adaptive-aperture one (in which the width of the kernel scales with the virial radius) and a fixed-aperture one  \citep[calculated with a fixed kernel, in a similar manner to][]{2015MNRAS.447.2683A,2015MNRAS.448.3665E}.
Our goal is to shed light on  the complex relation between the mass, web environment, and the fixed- and adaptive- aperture ambient densities of haloes, and in particular  address the issue of how these properties affect the abundance of haloes.

The paper is organized as follows. The methodology of the paper is presented in \S\ref{sec:methods} and the results are described in \S\ref{sec:results}. A general summary and discussion conclude the paper (\S\ref{sec:summary}).

%%%%%%%%%%%%%%%%%%%%%%%%%%%%%%%%%%%%%%%%%%%%%%
%%%%%%%%%%%%%%%%%%%%%%%%%%%%%%%%%%%%%%%%%%%%%%
\section{Methods}

\label{sec:methods}

The methodology employed here consists of extracting DM haloes from a DM--only $N$--body simulation by means of a halo finder. 
The environment within which a halo resides is defined by means of the cosmic web and by the local density field. 
The abundance of the haloes is expressed by mass and density functions. The aim of the paper is the study of the dependence of these functions on the mass, local density, and cosmic web environment of DM haloes.

\subsection{Simulation}
\label{subsec:Nbody}

The DM--only $N$--body simulation employed here has been performed as part of the Multimessenger Approach for Dark Matter Detection (MultiDark) project, and is dubbed the Small MultiDark Planck \citep[SMDP;]{2016MNRAS.tmp...48K} simulation. The cosmological parameters have the following values -  a cosmological constant density parameter $\Omega_\Lambda = 0.693$, a matter density parameter $\Omega_m = 0.307$, a Hubble constant of  $H_0 = 70$km s$^{-1}$Mpc$^{-1}$ , a spectral index of primordial density fluctuations given by $n_s = 0.96$, and a $\sigma_8 = 0.829$ normalization of the power spectrum. The simulation spans a box of side length $400\hmpc$ with $3840^3$ particles, achieving a mass resolution of $\sim 9.6 \times 10^7 h^{-1}M_{\odot}$.  DM haloes are defined by the BDM halo finder \citep{1997astro.ph.12217K, 2013AN....334..691R}, with their virial radius, $R_{vir}$, defined as the radius within which the density is equal to $360 \times \rho_m$, where $\rho_m$ is the average density in the universe.

%%%%%%%%%%%%%%%%%%%%%%%%%%%%%%%%%%%%%%%%%%%%%%
\subsection{Ambient density}
\label{subsec:density}

The dark matter density ($\rho$) field is evaluated on a $400^3$ Cartesian grid by the ``clouds in cells'' (CIC) algorithm, corresponding to a unit grid length of $1 \hmpc$. The density field is denoted here by $\Delta^{\rm CIC} = \rho / \bar{\rho}$, where $\bar{\rho}$ is the mean cosmological density. 
The raw CIC field is smoothed with a Gaussian kernel of width $R$, yielding
$\Delta{^{\rm CIC}_R}$. The smoothing length is taken here to be $R=1\hmpc$, a length scale larger than the virial radius of all but a small fraction of the most massive haloes in the simulation. \\
Note that the relative total volumes of the different web elements are purposefully left out of the density computations, as we are interested in calculating the different properties as they may appear to an observer.

%Fig.1
\begin{figure}
\includegraphics[width=20pc]{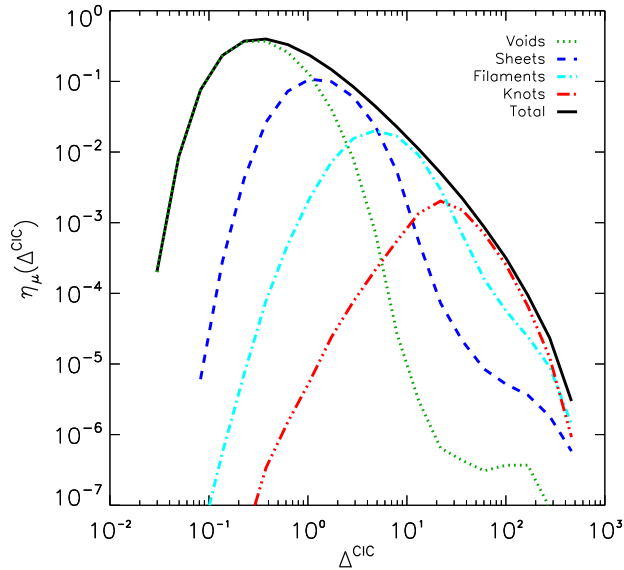}\hspace{0.cm}
\caption{The CIC density function, $\eta_\mu(\Delta{^{\rm CIC}_R}  )$, which measures the abundance of CIC cells of a given density $\Delta^{\rm CIC}_{R}$ that are classified as web type of $\mu=0$ (voids, green dotted line), $1$ (sheets, blue dashed line), $2$ (filaments, cyan dot-dashed line), and $3$ (knots, red dot-dot-dashed line). The black curve corresponds to the density function of all cells. This cosmic web element colour scheme is used throughout the paper. }
\label{fig:Dens_R_function}
\end{figure}

When dealing with the local density a halo resides in, namely its ambient density, one needs to ensure that the local density is smoothed on a scale larger than the halo's virial radius, \rvir, so as to reflect the halo's environment and not its internal structure. Two approaches are followed here - a fixed-aperture approach and a adaptive-aperture one. In the fixed-aperture approach the density field, $\Delta_R$, is the smoothed density, $\Delta{^{\rm CIC}_R}$ with $R=4\hmpc$, evaluated  at the center of the halo. 
The adaptive-aperture ambient density associated with a halo of mass $\mvir$, $\Delta_M$, is defined to be the mean value of $\Delta^{\rm CIC}$ within a top--hat sphere of $5$\rvir, where \rvir is the halo's virial radius (as defined in \ref{subsec:Nbody}). It follows that a  halo is endowed with two ambient densities, $\Delta_M$ and $\Delta_R$.

%%%%%%%%%%%%%%%%%%%%%%%%%%%%%%%%%%%%%%%%%%%%%%
\subsection{Cosmic web}
\label{subsec:Vweb}

%Fig.2
\begin{figure}
\includegraphics[width=20pc]{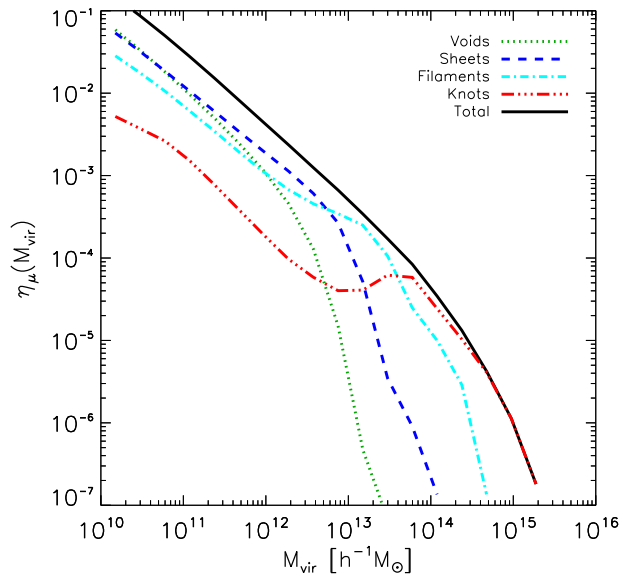}\hspace{0.cm}
\caption{The dependence of the mass function, $\eta_\mu(\mvir)$, on the cosmic web is plotted for all web types ($\mu=0,\  1,\  2,\  3$).  The well known relation between the mass function and the cosmic web is manifested here. }
\label{fig:mass_function}
\end{figure}

The normalized velocity shear tensor is defined at each grid  point as \citep{2012MNRAS.425.2049H}  :   
\begin{equation}
\Sigma_{\alpha\beta}=-\frac{1}{2H_0} \left( \frac{\partial v_{\alpha}}{\partial r_{\beta}} + \frac{\partial v_{\beta}}{\partial r_{\alpha}} \right), \;\;\; \alpha,\beta=x,y,z
\end{equation}
Here $H_0$ denotes Hubble's constant and the minus sign is introduced so as to associate positive eigenvalues with a converging flow. Spatial derivatives are calculated by means of the Fast Fourier  Transform (FFT) algorithm. Eigenvalues ($\lambda_i$, $i=1,2,3$) of the shear tensor are evaluated at each grid point, and assume the standard convention  of $\lambda_1 > \lambda_2 > \lambda_3$. The V--web is defined by introducing a threshold value, $\lambda_{\rm th}$, and by counting the number ($\mu$) of eigenvalues larger than $\lambda_{\rm th}$. It follows that $\mu=0, 1, 2, 3$ correspond to a V--web classification of voids, sheets, filaments, and knots, respectively \citep{2012MNRAS.425.2049H}. The V--web is defined by the (Gaussian) smoothing length used for calculating the shear tensor and by $\lambda_{\rm th}$.  A halo inherits its web type ($\mu$) from the web classification of the CIC cell where its center lies.  The V--web is introduced here as a measure of the environment of haloes, hence care needs to be taken to make sure that the web classification depicts the environment of a halo and not its internal structure, which is highly non-linear and where the web classification formalism does not apply. For that reason, the small fraction ($<0.01\%$) of haloes with \rvir\ greater than the basic smoothing length ($1\hmpc$) have their web classification drawn from a calculation of the cosmic web done over a density grid that is smoothed over $2\hmpc$, which is larger than the virial radius in the simulation. Following \citet{2012MNRAS.425.2049H} we adopt here a threshold value of $\lambda_{\rm th} = 0.45$.

%Fig. 3

\begin{figure*}
\includegraphics[width=20pc]{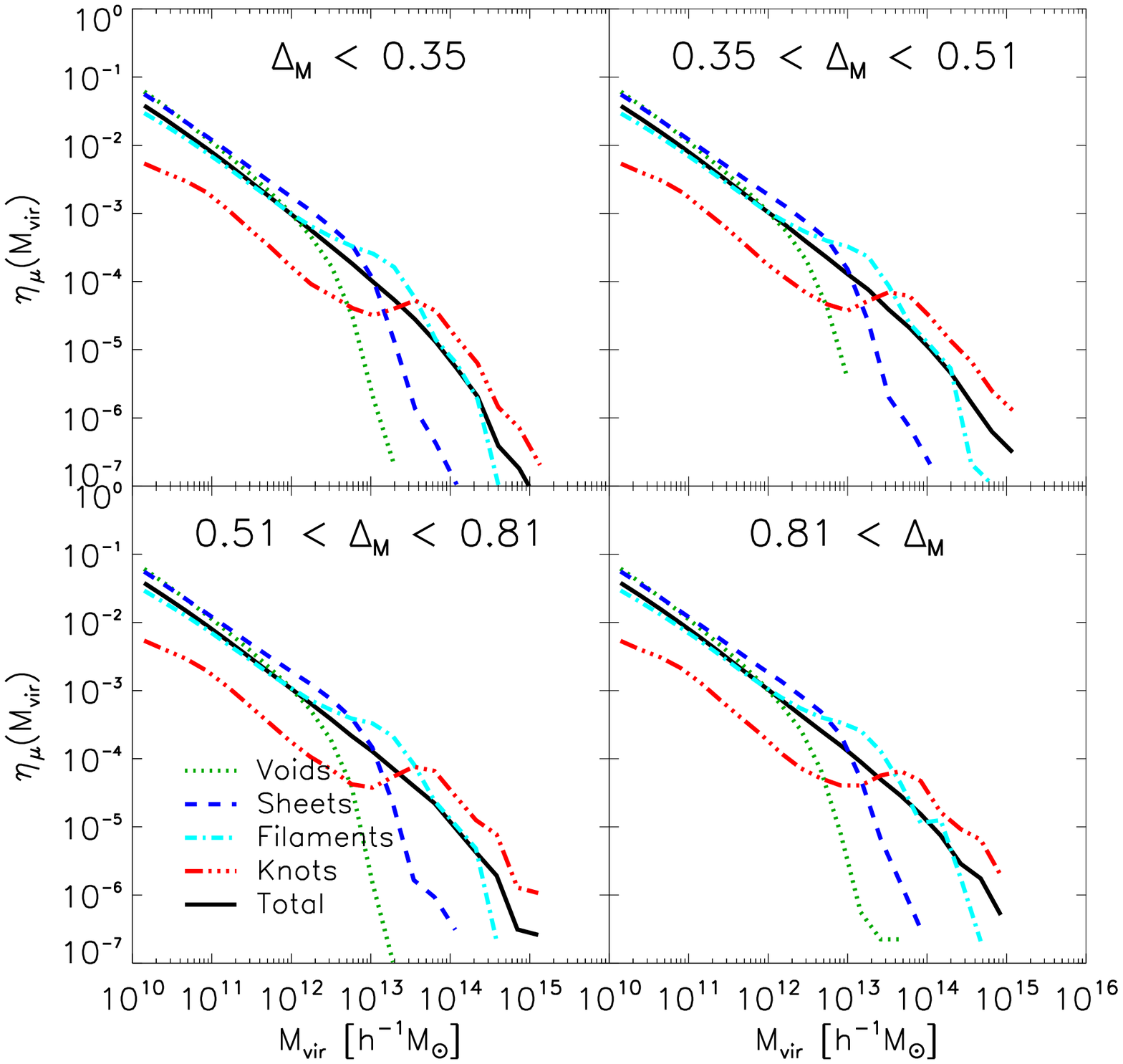}\hspace{0.cm}
\includegraphics[width=20pc]{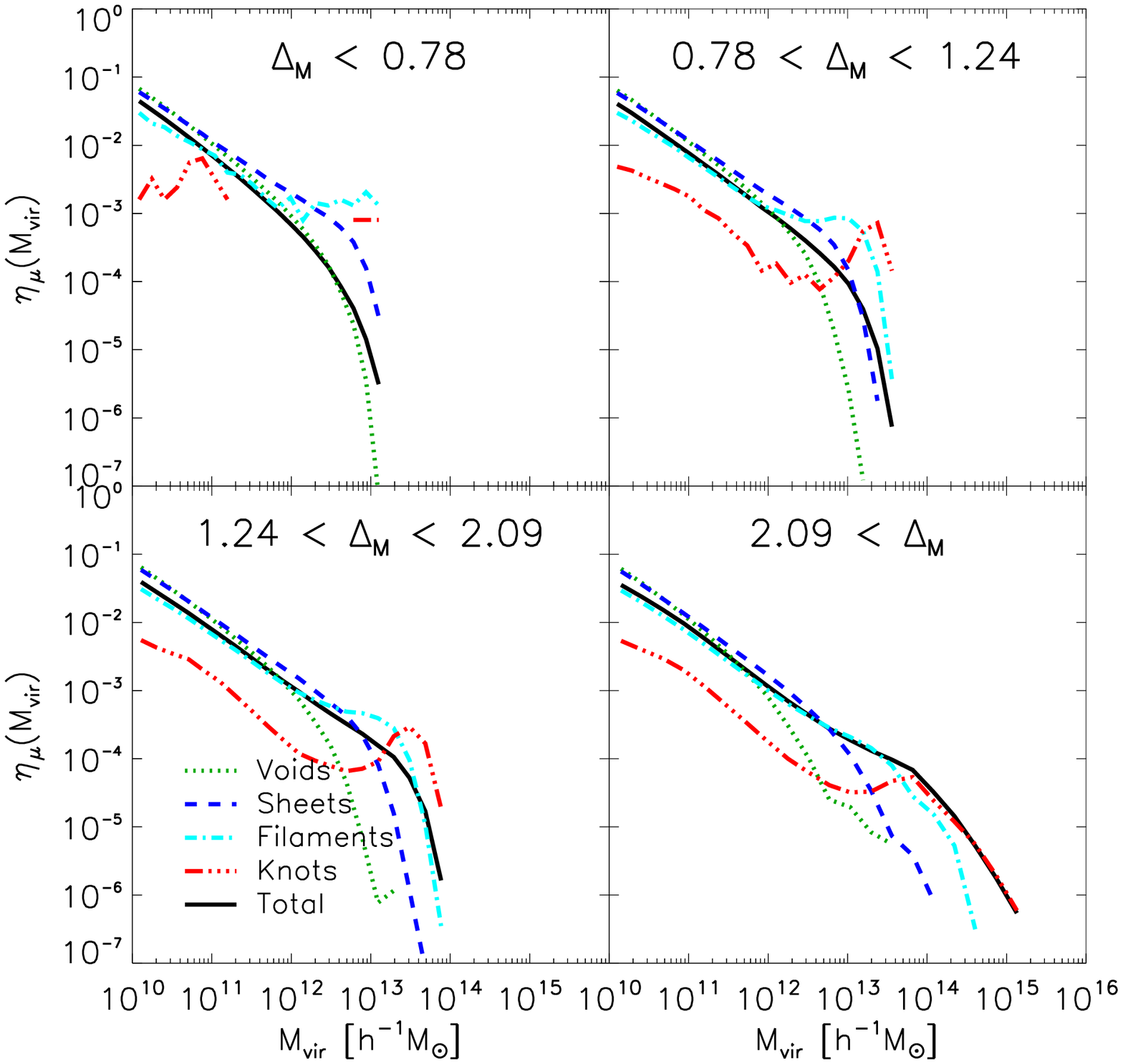}\hspace{0.cm}
\caption{The mass function conditioned by the ambient adaptive-aperture (left frame) and fixed-aperture (right frame) density of haloes is plotted for the different web elements, $\eta_\mu(\mvir \vert   \Delta_X)$ (where $X= M$ or $R$). 
}
\label{fig:eta_M_cond_Delta}
\end{figure*}

\subsection{Abundance of haloes: mass and density functions}
\label{subsec:mass_dens_functions}

DM haloes are characterized by their virial mass (\Mvir), ambient fixed- and  adaptive- aperture densities  ($\Delta_R$ and $\Delta_M$), and V--web type ($\mu$). Let  $n_\mu(\mvir,\Delta_X)$ be the total number of haloes per unit volume, as a function of the virial mass \Mvir and ambient adaptive- or fixed- aperture densities (with $X = M,R$), of a given web type $\mu$. Following \citet[and references therein]{2001MNRAS.321..372J} the function $\eta_\mu(\mvir,\Delta_X)$, which describes the abundance of haloes as a function of their mass, ambient (adaptive- or fixed- aperture) density, and web classification, is defined by:
\begin{equation}
\eta_\mu(\mvir,\Delta_X ) = \mvir \Delta_X \times {\d^2\  n_\mu(\mvir,\Delta_X) \over \d\ \mvir \  \d\ \Delta_X } 
\end{equation}
Conditional functions are defined here by:
\begin{equation}
\eta_\mu(\mvir \vert \Delta_X ) = \mvir \times {\d\  n_\mu(\mvir, {\rm within \ a \ given \ range \ of \ }\Delta_X) \over \d\ \mvir } 
\end{equation}
\begin{equation}
\eta_\mu(\Delta_X \vert \mvir ) = \Delta_X \times {\d\  n_\mu({\rm within \ a \ given \ range \ of \ } \mvir, \Delta_X ) \over \d\ \Delta_X  } 
\end{equation}
The mass and ambient density functions, $\eta_{\mu}(\mvir$) and $\eta_{\mu}(\Delta_X	)$, are obtained by marginalizing over the other parameter.

\section{Results}
\label{sec:results}

\begin{table*}
\begin{tabular}{ccccc}
\hline
Function                                             &  $M$  &  web type ($\mu$)  &  $\Delta_R$  &  $\Delta_M$  \\
\hline
$\eta(\mvir\vert \mu)$ (Fig. 2)                                          & +  & + &      &     \\
$\eta_\mu(\mvir \vert \Delta_M)$ (Figs. 3 \& 4)                        & +  & + &      &  -  \\
$\eta_\mu(\mvir \vert \Delta_R)$ (Figs. 3 \& 4)                        & +  &  + &  +  &    \\
$\eta_\mu(\Delta_M \vert M_{vir})$ (Fig. 5)                                & -  & +  &      &  + \\
$\eta_\mu(\Delta_R \vert M_{vir} )$ (Fig. 5)                                &  +  & + &   + &     \\
\hline
\end{tabular}
 \caption{A summary of the dependence of the various conditional functions on the four attributes of the DM haloes: mass ($M$), web classification ($\mu$), fixed-aperture ($\Delta_R$) and adaptive-aperture ($\Delta_M$) ambient densities.}
 \label{table1}

\end{table*}

%Fig. 4

\begin{figure*}
\includegraphics[width=20pc]{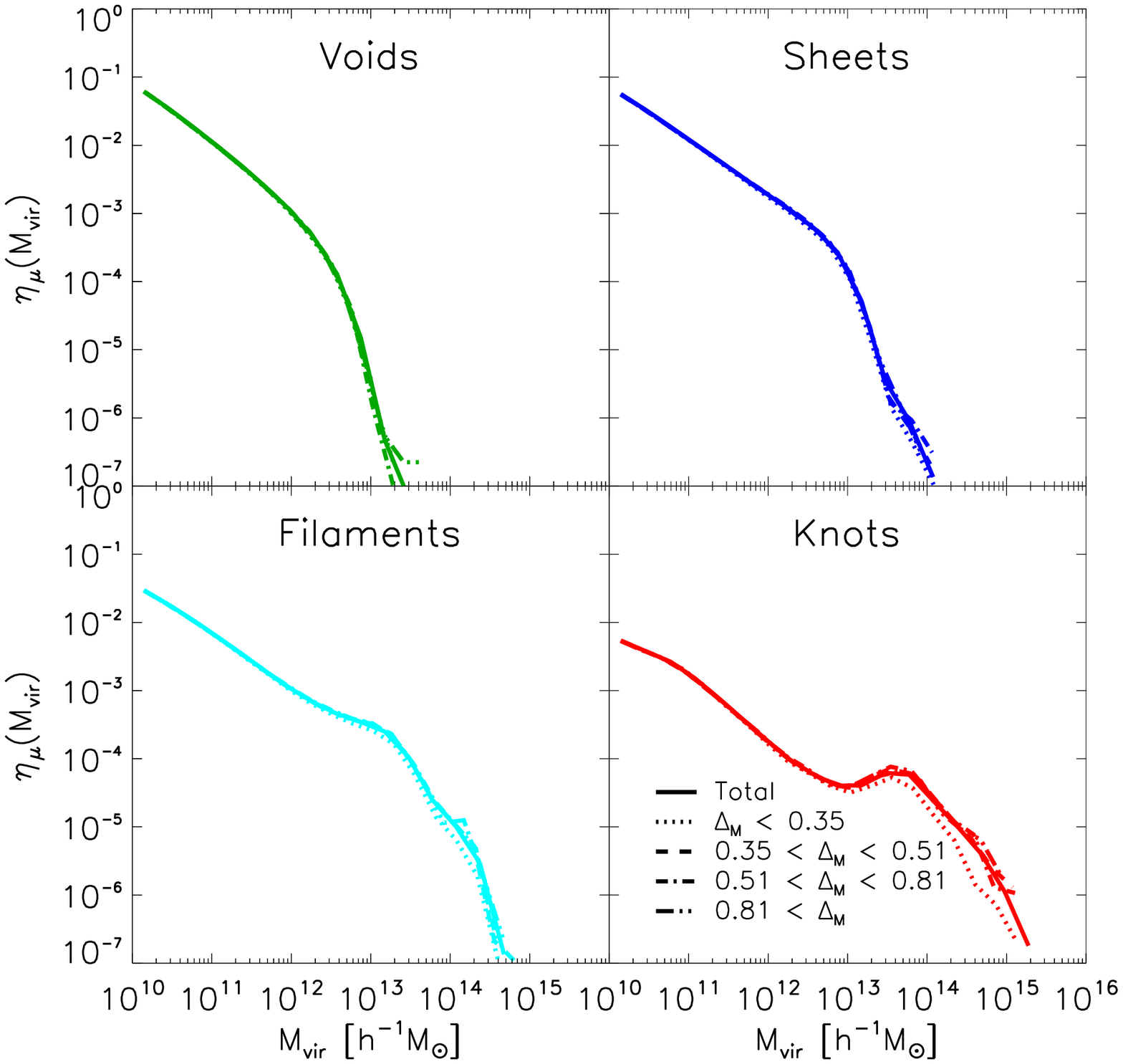}\hspace{0.cm}
\includegraphics[width=20pc]{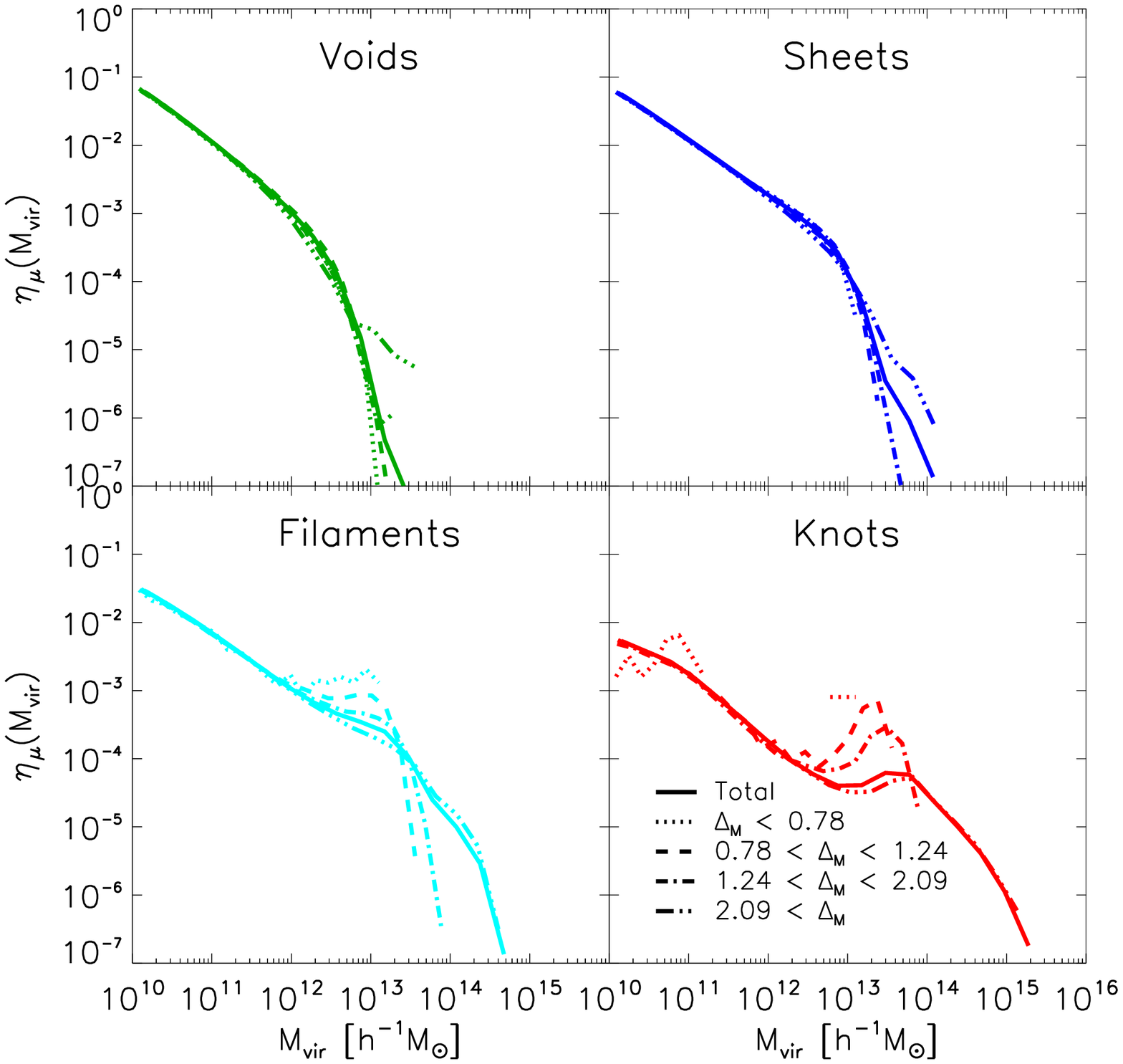}\hspace{0.cm}
\caption{The mass function conditioned by the four web types ($\eta_\mu(\mvir \vert \Delta_X)$, where $X=M$ (left) and $R$ (right): Each frame shows the mass function  of a given web type, with the different curves corresponding to different adaptive-aperture (left frame) and fixed-aperture (right frame) ambient density ranges. Here the line styles correspond to different density ranges (given as legends in the plots), with the solid line depicting the mass function of all haloes in this web element, regardless of ambient density.}
\label{fig:eta_M_cond_mu}
\end{figure*}

Our analysis commences with the evaluation of the abundance of CIC cells with Gaussian smoothed density $\Delta{^{\rm CIC}_R}$ as a function of the web attribute $\mu$. Fig. \ref{fig:Dens_R_function} presents the CIC density function,
$\eta_\mu(\Delta{^{\rm CIC}_R}) = \d\  n_\mu(\Delta{^{\rm CIC}_R}) /   \d\ln\Delta{^{\rm CIC}_R}$. While there is a significant overlap in the density distributions of the different web elements, the plot reproduces the well known correlation between the cosmic web classification and the density \citep[and references therein]{2012MNRAS.425.2049H}. The voids are populated predominantly by low density cells and the transition from voids, through sheets and filaments, to knots, corresponds to a systematic shift from low to high density of the CIC cells.
 
Fig. \ref{fig:mass_function} reproduces another  well known  relation, namely the correlation between the mass of haloes and their  web environment \citep[and references therein]{2014MNRAS.441.2923C,2015MNRAS.446.1458M}. It shows the mass function, $\eta_\mu(\mvir)$, for the four  web environments. The tendency of low mass haloes to reside in voids, and 
the more massive ones to be found, progressively, in sheets, filaments and knots, is clearly visible here.

The analysis   shifts here to include  the effects of the ambient density. Fig. \ref{fig:eta_M_cond_Delta} presents the conditional, by the adaptive-aperture (left frame) and fixed-aperture (right  frame) ambient density, mass function.  Each panel (i.e. quadrant of the frame) presents the mass function of haloes of a given ambient density interval as a function of their mass and web attribute. 
Bins of    $\Delta_M$ and $\Delta_R$ have been selected to have an equal number of haloes in each bin. The figure clearly shows that at a given ambient density cut the mass function varies significantly  with web classification. This holds for both  the adaptive- and fixed- aperture densities. 

Fig. \ref{fig:eta_M_cond_mu} presents again the conditional mass function, but the role of the ambient density (adaptive-aperture, left frame, and fixed-aperture,  right frame) and the web attribute, $\mu$, is reversed. Each panel corresponds to a given web attribute (voids, sheets, etc.) and the different curves correspond to different intervals of the ambient density. The curves here are essentially the same as those shown in Fig.~ \ref{fig:eta_M_cond_Delta}, yet grouped and presented differently for clarity. The left frame of Fig. \ref{fig:eta_M_cond_mu} shows  that for a given web environment the variation of the adaptive-aperture ambient density does not change the mass function.  A modest variation is found for the fixed-aperture case for high mass knot and filament haloes.

\begin{figure*}
\includegraphics[width=20pc]{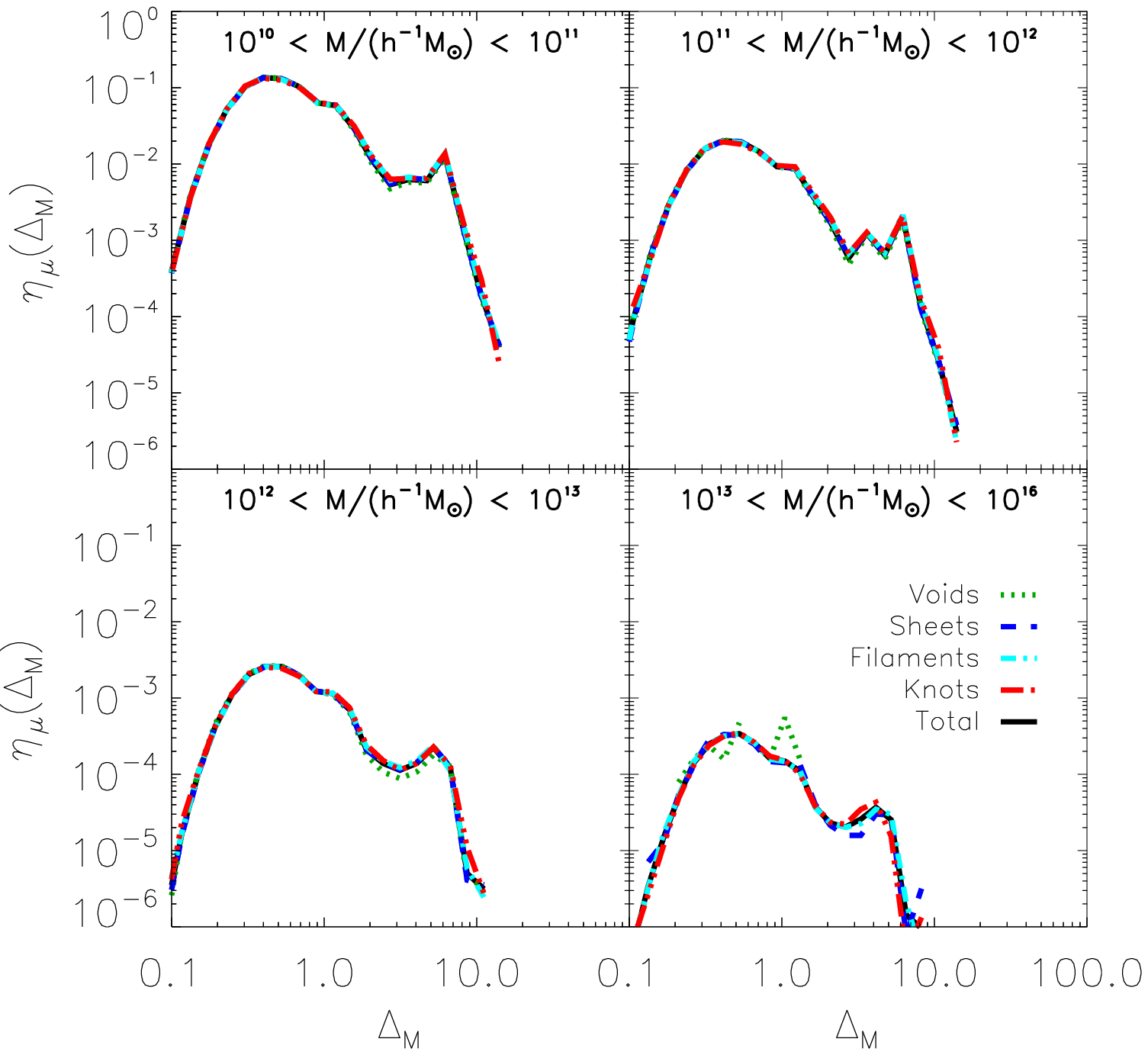}\hspace{0.cm}
\includegraphics[width=20pc]{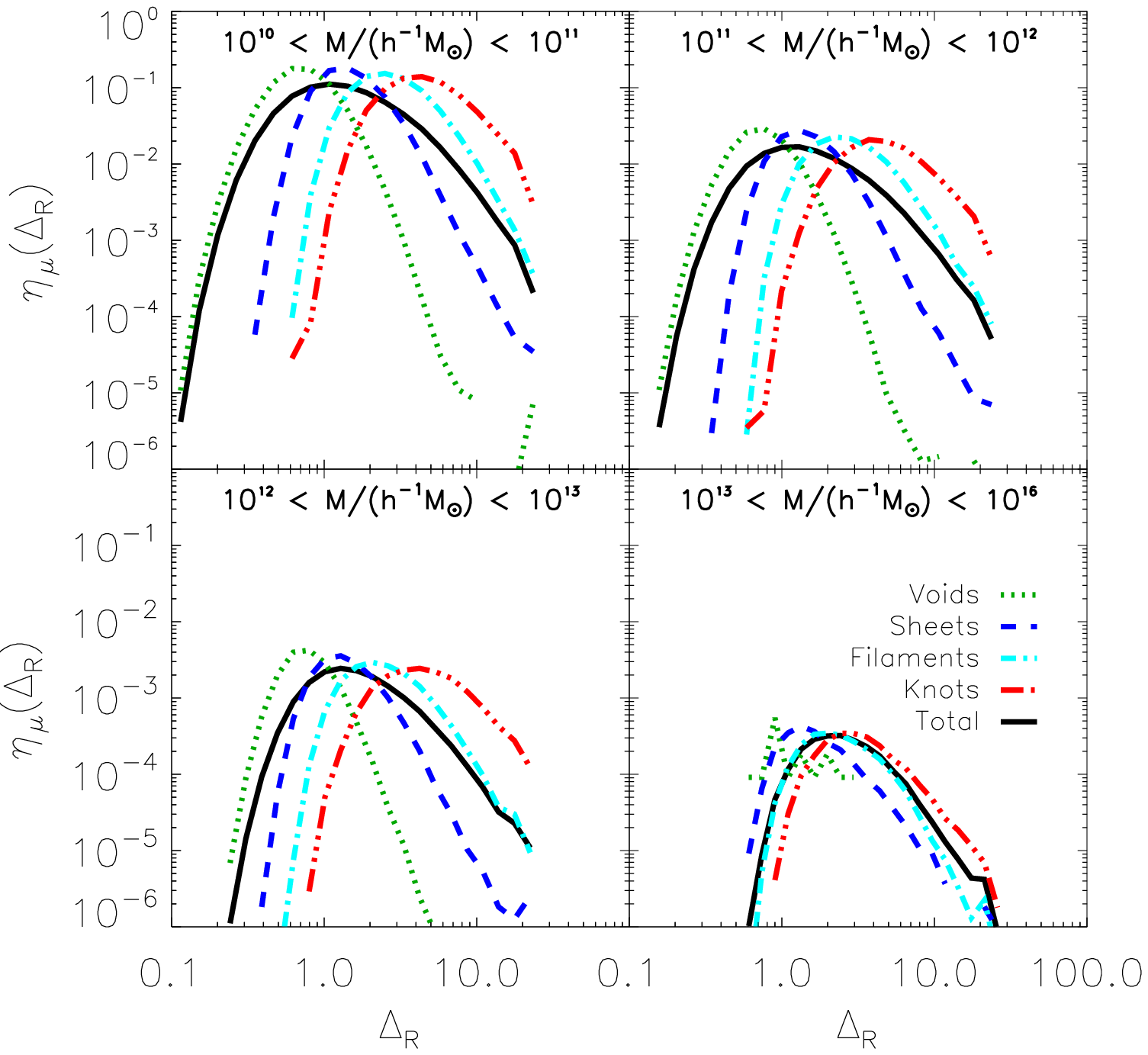}\hspace{0.cm}
\caption{The adaptive-aperture (left frame) and fixed-aperture (right frame) ambient density functions, $\eta_\mu(\Delta_X \vert M)$ (where $X= M,R$), conditioned by the virial mass of haloes, is plotted for the different web elements. 
}
\label{fig:eta_Delta_cond_mu}
\end{figure*}

Next, the conditional ambient density function, namely $\eta_\mu(\Delta_X \vert M_{vir})$ (where $X=M,R$), is considered. Fig. \ref{fig:eta_Delta_cond_mu} depicts the adaptive-aperture (left frame) and fixed-aperture (right frame) ambient density functions. Individual panels show    $\eta_\mu(\Delta_X \vert M_{vir})$ at a given mass range   for the full range of web types.
The $\eta_\mu(\Delta_M \vert M_{vir})$ curves, at a given mass interval, are  virtually independent of the web type. Namely, the number of haloes as a function of their adaptive-aperture ambient density, at a given mass range, does not vary with the web environment.  This is not the case for the fixed-aperture ambient density (right panel of Fig.~\ref{fig:eta_Delta_cond_mu}), where the mass function shows a significant dependence on the environment.  

Figs. \ref{fig:eta_M_cond_mu} and \ref{fig:eta_Delta_cond_mu} manifest a decoupling between adaptive-aperture ambient density and the web environment of haloes. The number of haloes as a function of their mass depends on the web environment but is independent of $\Delta_M$. The number of haloes as a function of $\Delta_M$ depends on the mass but is independent of the environment. These trends are not as strong in the fixed-aperture ambient density case, where some difference is clearly seen between the curves. A summary of the dependence, or lack of it, of the various functions is presented in Table \ref{table1}.

%Fig. 5

\begin{figure*}
\includegraphics[width=20pc]{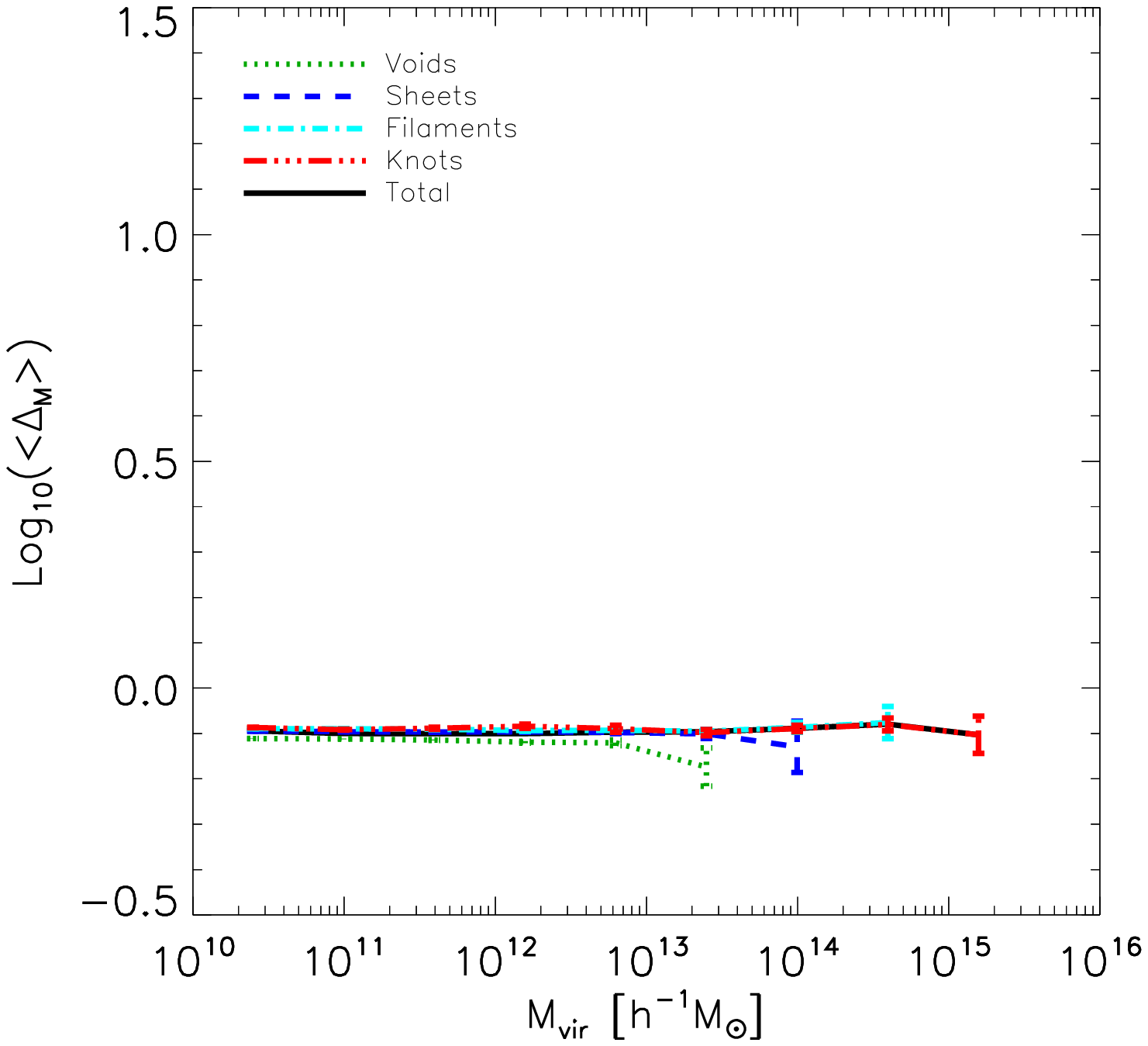}
\hspace{0.cm}
\includegraphics[width=20pc]{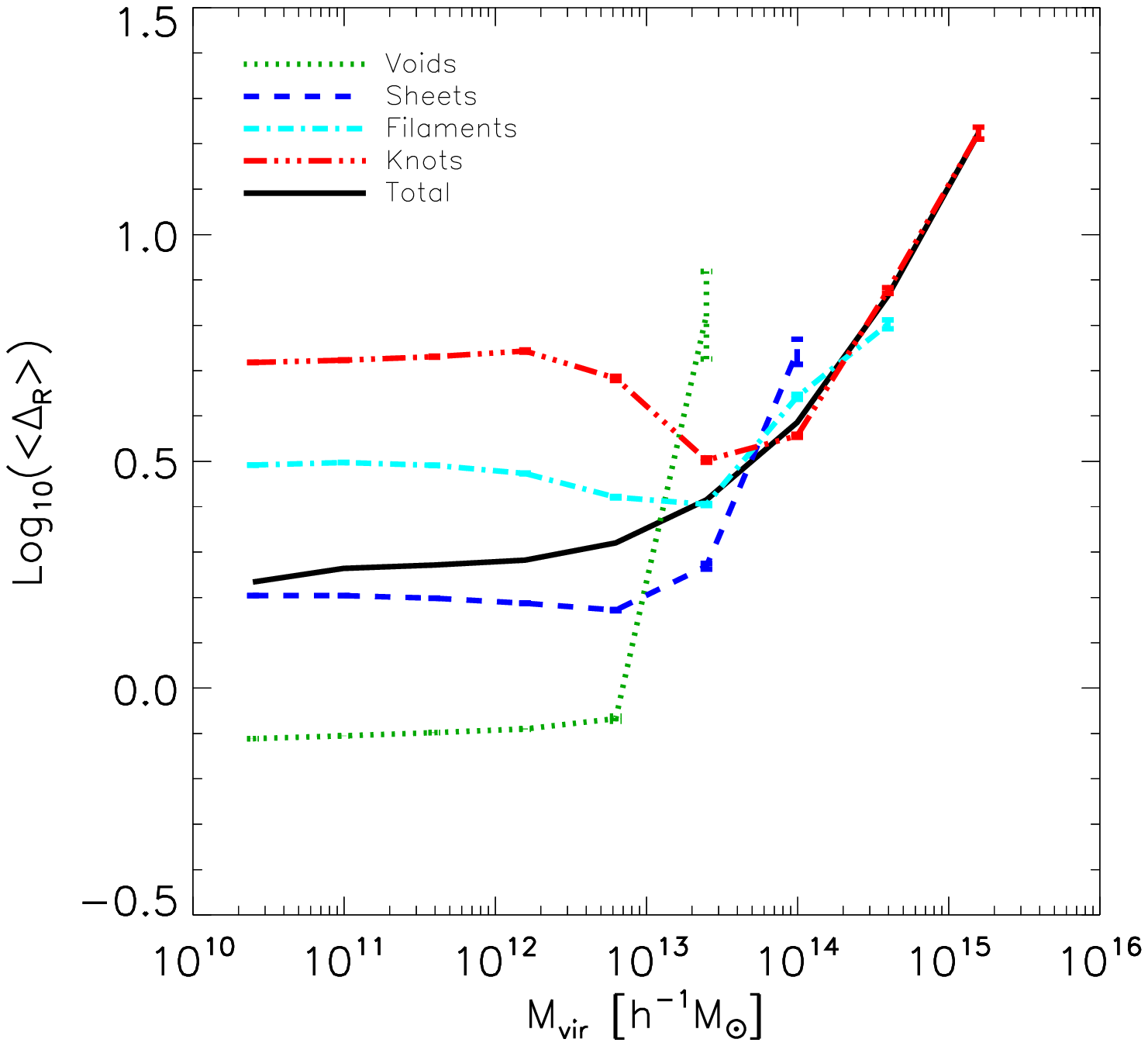}
\hspace{0.cm}
\caption{The mean adaptive-aperture (left frame) and fixed-aperture (right frame) ambient density of haloes plotted against the virial  mass, for the different web environment.
}
\label{fig:mean_density}
\end{figure*}

\begin{figure*}
\includegraphics[width=20pc]{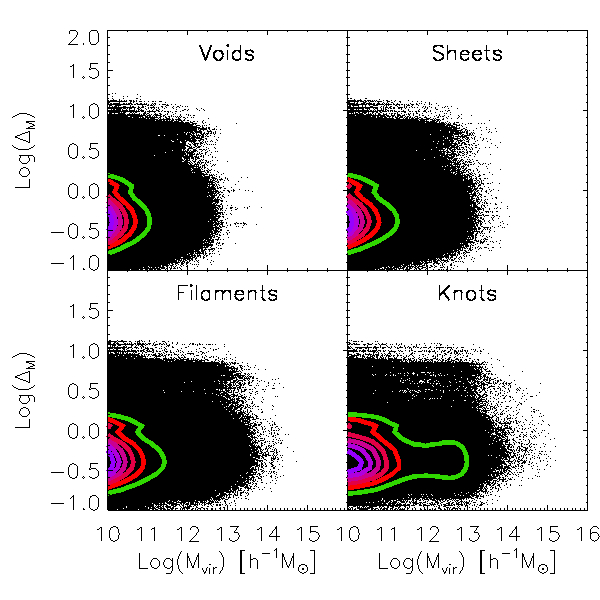}\hspace{0.cm}
\includegraphics[width=20pc]{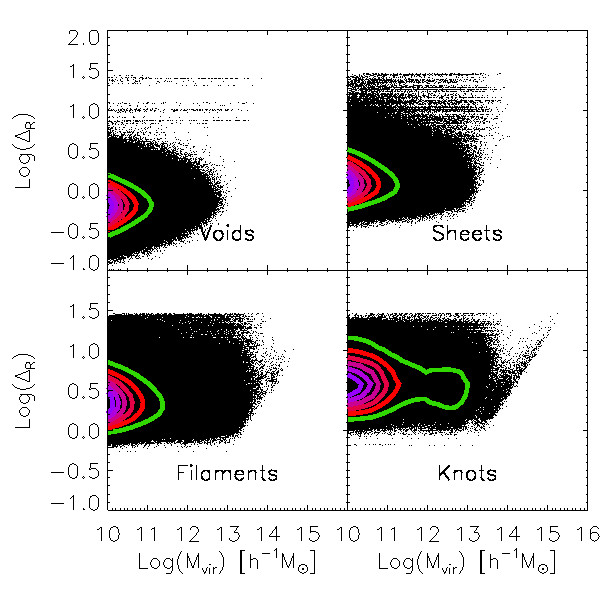}\hspace{0.cm}
\caption{Scatter plots of the haloes in the parameter space of their ambient adaptive-aperture (left frame) and fixed-aperture (right frame) density and their virial mass.  The coloured contours indicate the number density of haloes in that parameter space, and are present in order to point out the shape of the distributions; the actual numerical values are of no significance to our results.}
\label{fig:contours}
\end{figure*}

Recall that the main difference between the adaptive- and fixed- aperture ambient densities is that the former is calculated within a radius that scales with the virial radius of a halo while the latter is calculated within a fixed radius. It follows that a trivial correlation is induced between the mass and $\Delta_R$ - a sphere of a radius of $R=4\hmpc$ that contains a $10^{14}\hmsun$ halo is (statistically) bound to have a higher density than the one that contains a  $10^{11}\hmsun$ halo. This dependence introduces a leakage of the mass dependence of the mass function to a dependence on  $\Delta_R$. Such a dependence is not expected for the adaptive-aperture ambient density for haloes of mass below $M_\ast$, the characteristic  mass of the mass function, beyond which the self--similarity of the mass function breaks. This is clearly illustrated by Figs. \ref{fig:mean_density} and \ref{fig:contours}. Fig~\ref{fig:mean_density} presents the mean adaptive-aperture (left frame) and fixed-aperture (right frame) ambient densities as a function of the virial mass for the different web environments. The mean values of the adaptive-aperture ambient density remain constant. The fixed-aperture ambient density, on the other hand,  shows a clear departure from this constancy, and is changing both with halo mass and with web environment.
Fig. \ref{fig:contours} shows the full distribution of the haloes in the $M_{vir} - \Delta_X$ ($X=R,M$), plotted separately for each web environment. The distribution is shown by means of a scatter and   contours plots. The figure clearly shows the virtual invariance of the distribution along the vertical (ambient density) axis for $\Delta_M$,   while for $\Delta_R$ the distribution depends both on the virial mass  and the web environments.

\section{Summary and discussion}
\label{sec:summary}

Dark matter haloes serve as the building blocks of the large scale structure of the universe, and a very crude and biased proxy for galaxies. An open question in cosmology and the theory of structure formation is to what extent does environment affect the properties of galaxies and haloes. The present paper aims at shedding light on the problem from a structure, rather than galaxy formation, standpoint. The paper focuses on the analysis of DM--only simulations and addresses the issue of how the environment affects the abundance of DM haloes. Haloes are assigned  four attributes: their virial mass, their fixed- and adaptive-aperture ambient densities  (termed $\Delta_R$ and $\Delta_M$ and defined as the Gaussian smoothed  density within 4$\hmpc$ and the top--hat smoothed within 5 virial radii, respectively), and their cosmic web classification (i.e. voids, sheets, filaments, and knots)  according to the eigenvalues of the velocity shear tensor. The main result of the paper is that the mass function of haloes, namely the abundance of haloes as a function of their mass, depends on the haloes' web environment and not on their adaptive-aperture ambient density. The weak dependence on the fixed-aperture ambient density is induced by the cross correlation between the mass of a halo and its fixed-aperture ambient density.

Is this result in conflict with \citet{2015MNRAS.447.2683A}, and their statement on the dependence of the halo abundance on the ambient density? Technically speaking, no. Those authors used the fixed-aperture method to calculate the densities, and evaluated the tidal tensor with a the same fixed smoothing kernel; they also took into account the difference in the total mass of each web element when calculating their mass functions. This makes comparison with our results problematic. When a adaptive-aperture ambient density is used, the dependence of the mass function on the ambient density vanishes, leaving the sole dependence to be on the web environment.

%Fig. 6

The present work, together with \citet{2015MNRAS.446.1458M}, leads to a coherent picture regarding how galaxies are affected by their neighbourhood.  This earlier work, in which hydrodynamical simulations of galaxy formation were studied, suggested that the halo mass is the dominant parameter that shapes the fate of  the baryons in their parent haloes. Here we show that the abundance of haloes as a function of their mass is closely correlated with their web environment, leaving no room for further dependence on their (adaptiv-aperture) ambient density. So it is this strong mass--web dependence that drives the apparent dependence of the galaxy properties on the environment, and it is an environment defined by the cosmic web, and not by the (adaptiv-aperture) ambient density.

\begin{center}
\textbf{Acknowledgments}
\end{center}

NIL is supported by the Deutsche Forschungs Gemeinschaft. \\
YH has been partially supported by the Israel Science Foundation (1013/12). \\
The authors would also like to thank Dr. David Alonso for helpful explanations and Prof. John Peacock for a very insightful report.

%Fig. 7

\setlength{\bibhang}{2.0em}
\setlength\labelwidth{0.0em}
\bibliography{./ref}
\end{document}